# Anomalous frequency scaling of acoustic phonon damping in nickel cavities fabricated by ps-laser delamination

Alba Viejo Rodríguez[1†], Andrea Rossetti[1,9†*], Marco Gandolfi[2,3,4†], Yoav Urbina Elgueta[5], Evgeny B. Modin[5], Svetlana Starikovskaia[6], Tat Loon Chng[6], Vasily Temnov[6], Maria Antonietta Vincenti[2,3,4], Daniele Brida[1], Paolo Vavassori[5,7], and Nicolò Maccaferri[8*]

[1]Department of Physics and Materials Science, University of Luxembourg, 162a Avenue de la Faïencerie, 1511 Luxembourg, Luxembourg

[2]Dipartimento di Ingegneria dell'Informazione, Università degli Studi di Brescia, Via Branze 38, 25123 Brescia, Italy

[3]Consiglio Nazionale delle Rircerche - Istituto Nazionale di Ottica, Via Branze 45, 25123 Brescia, Italy

[4]Consorzio Nazionale Interuniversitario per le Telecomunicazioni (CNIT), Viale G.P. Usberti 181/A Sede Scientifica di Ingegneria-Palazzina 3, 43124 Parma, Italy

[5]CIC nanoGUNE, Tolosa Hiribidea, 76, 20018 Donostia-San Sebastián, Spain

[6]LSI, Ecole Polytechnique, CEA/DRF/IRAMIS, CNRS, Institut Polytechnique de Paris, 91128 Palaiseau, France

[7]IKERBASQUE Basque Foundation for Science, Plaza Euskadi 5, 48009 Bilbao, Spain

[8]Department of Physics, Umeå University, Linnaeus väg 24, 901 87 Umeå, Sweden

[9]Department of Physical Chemistry, Fritz Haber Institute of the Max Planck Society, 14195 Berlin, Germany

*rossetti@fhi-berlin.mpg.de

*nicolo.maccaferri@umu.se

†These authors contributed equally

## ABSTRACT

Single-shot picosecond (ps) laser induced delamination allows for the direct generation of suspended membranes from a continuous metallic film, offering a promising platform for control of ultrafast magnetization dynamics driven by acoustic waves. Using the picosecond-ultrasonics method, we demonstrate that long-lived low-frequency acoustic waves can be optically-excited in the delaminated cavities. At the same time, higher-frequency modes >60GHz exhibit a surprisingly fast damping, following a scaling law incompatible with the expected attenuation mediated by phonon-phonon scattering. Comparing measurements between delaminated cavities and a benchmark nickel film in contact with the substrate, we link our findings with structural modifications of the nickel crystal induced by the delamination process.

Opto-acoustic properties of materials[1,2] can be exploited in a broad range of applications, for instance to modulate light intensity[3,4], control magnetization dynamics[5,6], gravitational wave detection and quantum opto-mechanics[7], as well as ultrafast and all-optical switching[8,9] and ultrasound detection techniques[10] for the characterization of nanoscopic features in materials[11,12]. Furthermore, elastic strain engineering uses stress to boost material properties[13], including enhanced electron mobility in semiconductors for more efficient photovoltaic devices[14] and transistors[15]. In mechanical engineering, the pursuit of resonators with high quality factors is pivotal for a broad range of applications, from ultralow noise opto-mechanical quantum technologies[16] to energy-efficient magneto-acoustics in spintronic devices[17,18]. Notably, coupling between acoustic and magnetic degrees of freedom in thin-film cavities can lead to resonant enhancement of magneto-optical effects[19]. These perspectives motivate the search for optimized nano-resonators with lower damping and enhanced propagation of acoustic waves.

In practice, removing the substrate below a thin film can provide high-quality resonators, since the energy of the acoustic pulse is physically trapped within the film due to impedance mismatch[20,37]. In a seminal work by Kim et al. [6], picosecond acoustic pulses trapped within a freestanding nickel film were used to control magnetization precession on ultrafast timescales. Despite this, opto-acoustic control of magnetization dynamics has not yet moved past the proof-of-concept stage, due to complex methodologies needed to fabricate the freestanding film. For example, in ref. 6 the freestanding film was realized by first growing a sacrificial salt layer on top of a $SiO_2$ substrate. A thin nickel film was then grown on top and separated from the substrate by dissolving the salt layer in water. The freestanding film was then stretched and clamped onto a holder with a center hole. A simpler methodology could help realizing the technological potential of ultrafast magnetization control into practical applications.

In this context, laser delamination was recently proposed as a way to deterministically pattern cavities detached from the substrate at only a fraction of usual time and fabrication costs[21,22]. It was shown that, carefully tuning the laser fluence, micrometric regions where the nickel film detaches from the substrates are created[22]. Once properly controlled, this process promises direct on-demand patterning of suspended membranes on areas down to the microscale and supporting a unique combination of multiple excitations (magnetic, optical, acoustic)[19]. So far, existing works have focused on topographical characterization of nickel cavities fabricated via laser delamination (from now on referred to as ‘delaminated cavities’), employing optical interferometry and electronic microscopy techniques. Currently, an experimental study of the opto-acoustic response of delaminated nickel cavities is lacking. In the different scenario of ref. 6 a surprisingly high attenuation of the acoustic pulses was observed in the freestanding film compared to a benchmark sample that did not undergo the substrate removal procedure. In the context of this paper, the complex thermo-mechanical dynamics[21] leading to the re-solidification of the nickel crystal in proximity of the detached region might affect acoustic-pulse propagation in different ways after the fabrication of the delaminated cavities. Overall, alterations of the crystalline structure and of the interface quality could reshape the opto-acoustic response of our structures beyond the targeted reduction of damping directly associated to the absence of a substrate. Understanding the mechanisms governing the propagation of optically-induced picosecond acoustic pulses in delaminated cavities therefore represents a necessary next step to assess possible integration in actual devices.

In this work we characterize the opto-acoustic response of large (>10 μm) micrometric bubbles of freestanding nickel films fabricated via laser delamination. The freestanding nickel cavities were fabricated using a pulsed laser focused on 230 nm thick nickel thin film evaporated on top of a $SiO_2$ substrate. For a laser fluence slightly (a few percent) below the ablation threshold it is possible to obtain the formation of closed delamination cavities[22] (more details on the fabrication can be found in Supplementary Note 1). We benchmark our findings with the better-known scenario of a nickel thin film in contact with a $SiO_2$ substrate[5] (from now on referred to as 'pristine film'). Scanning electron microscope (SEM) images of both the pristine film and a typical delaminated cavity produced by our method are presented in Figure 1a,b (details on sample preparation for SEM imaging can be found in Supplementary Note 2). As shown in Figure 1b, the ps-laser delamination induces the coalescence of small crystal grains of the pristine film into large crystallites in the delaminated cavity, accompanied by a significant increase of roughness at the nickel/air bottom interface.

We use the pump-probe method to investigate the propagation of picosecond acoustic pulses within the delaminated cavity. Comparing these findings with benchmark measurements on the pristine nickel film, we can understand how the ps-laser delamination process impacts the opto-acoustic performance of our structures. Additional information on the experimental setup can be found in Supplementary Note 3. A sketch of the interaction between pump beam and sample is shown in figure 1c. The pump beam, whose penetration is shorter (roughly 30nm, corresponding to the optical skin depth of nickel at 515nm) than the thickness of both delaminated cavity and pristine film, heats up the upper part of the system (shaded red region in figure 1c). This localized heating triggers longitudinal acoustic waves that travel towards the lower part of the nickel domain. After reaching the lower nickel/air (delaminated cavity) or nickel/$SiO_2$ (pristine film) boundary, the acoustic pulse is partially reflected back to the first interface, where the probe pulse detects the effect of the strain on the transient optical response of the system. The periodic modulation of the refractive index due to the travelling acoustic wave shapes the pump-probe traces with a series of "echoes". In the case of the film, we expect that the energy of the acoustic pulse is gradually dissipated due to transmission losses at the bottom nickel/$SiO_2$ interface. This effect should be significantly suppressed in the case of the delaminated cavity, due to the higher impedance mismatch at the nickel/air bottom interface. However, the larger crystallites and higher roughness of the bottom nickel interface caused by the delamination process can enhance other loss mechanisms, for example scattering of the acoustic pulse from defects.

Analyzing the pump-probe acoustic echoes in the time and Fourier domain we can understand which mechanisms govern the time-evolution of the acoustic pulse in the delaminated cavity. In Figure 2a we show the transient reflectance curves measured under identical excitation conditions (fluence equal to 5 $mJ/cm^2$) for five different delaminated cavities and two different spots on the pristine film. To better isolate the opto-acoustic response, an exponentially decaying baseline arising from electron-electron and electron-phonon scattering is subtracted from all pump-probe traces (see Supplementary Figure S4a).

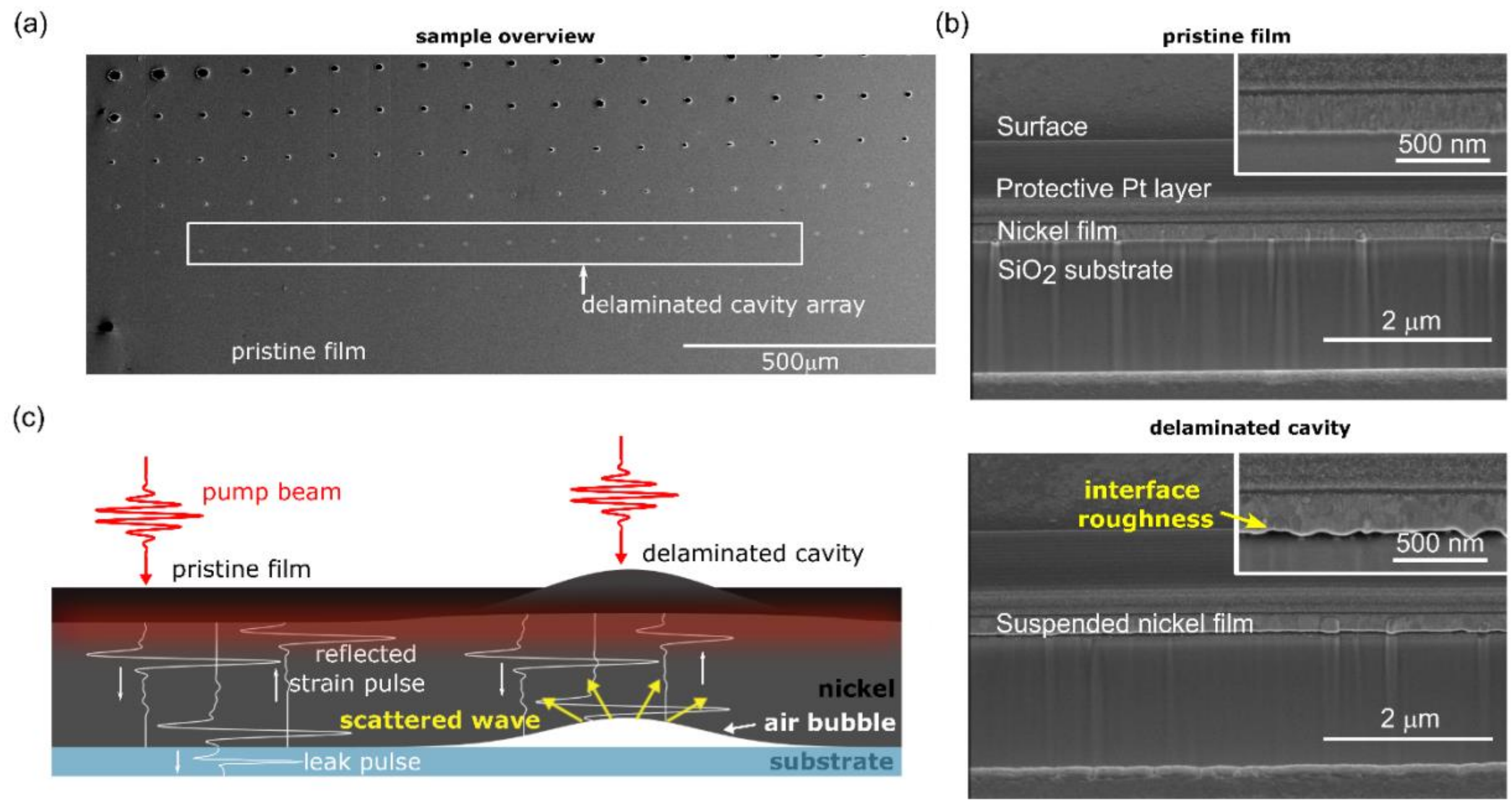

**Figure 1.** (a) SEM overview of the patterned Ni film investigated in this work. The white rectangle highlights an array of delaminated cavities. For comparison, the circular structures in the top rows (black circles) are regions where the nickel is completely removed from the substrate (obtained for fluence values exceeding the ablation threshold). (b) SEM cross-section cuts of the pristine nickel film (top) and delaminated cavity (bottom). Insets: magnified image of the pristine film and delaminated cavity cross-section. The yellow arrow highlights the corrugations of the bottom nickel interface induced by the delamination process. The protective Pt layer was used only for imaging purposes and not in the cavity that were measured. (c) Sketch of the excitation of an acoustic pulse on the pristine film/delaminated cavity, after interaction with the pump pulse.

The periodicity of the acoustic echoes depends on the thickness of the cavity/film and the sound velocity in nickel. Considering a thickness of 230 nm, and a sound velocity of longitudinal acoustic waves of 6.04 nm/ps[23], we can expect a periodicity for the roundtrip of 76 ps, corresponding to a fundamental acoustic frequency of approximately 13GHz. This number roughly matches the observed periodicity in the measured data for both pristine film and delaminated cavities (see Figure 2a). The shape of the acoustic echoes depends on the relative amplitudes and phases of the different acoustic eigenmodes propagating in the cavity. While the curves measured on different spots of the pristine film are practically identical (see two top rows of Figure 2a), we observe a different response between delaminated cavities, plausibly due to variations in the corrugation profile of the bottom interface.

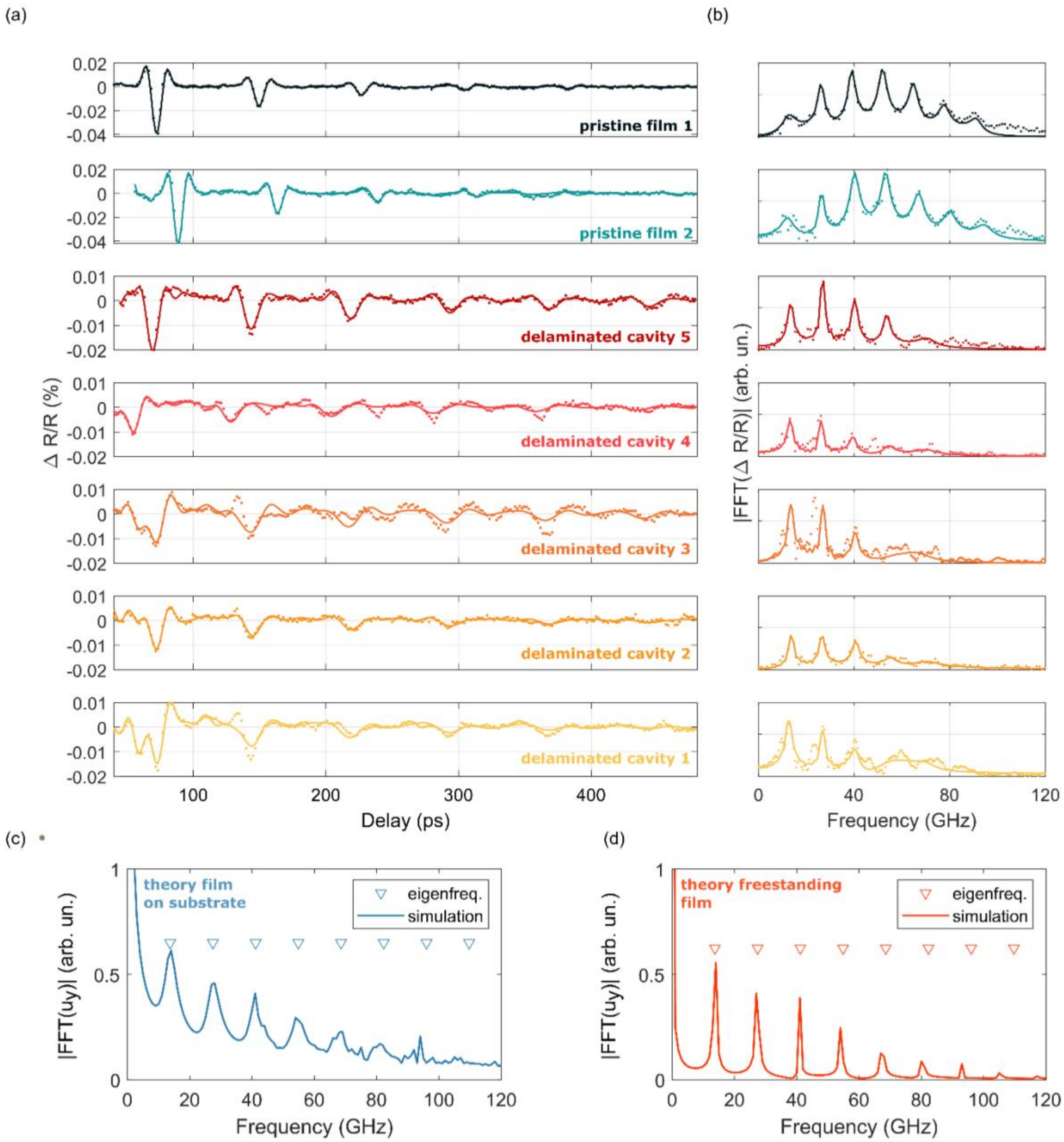

**Figure 2.** (a) Transient reflectivity measurements (dots) and fit (solid line) for two spots on the pristine nickel film and five different delaminated cavities. (b) Fast-Fourier-transform of the curves in a. (c) Spectral position of the expected eigenfrequencies (triangles) and FFT of the simulated displacement field $u_y$ for a thin film on top of a $SiO_2$ substrate with thickness identical to the nominal value of our samples. (d) same as (c), but for a freestanding film.

Compared to the pristine film, we can in general appreciate a smaller attenuation of the acoustic pulse between successive roundtrips in the delaminated cavities, accompanied by a distortion of the acoustic echoes. For example, in delaminated cavity 3 the initial acoustic pulse transforms into an almost sinusoidal oscillations with a period of roughly 38ps, indicating that after three roundtrips the response is dominated by a mode at twice the fundamental acoustic frequency. These observations suggest that the various eigenmodes experience different reflection conditions at the bottom interface and that a mechanism more complex than a simple reduction of damping governs the overall opto-acoustic response of the delaminated cavities.

To better understand acoustic-wave propagation in our target and benchmark samples, in Figures 2b we show a Fourier transform of the experimental transient reflectance curves. In all cases we observe a series of peaks, each one corresponding to a different eigenmode contributing to the overall acoustic wave-packet propagating in the two systems. We can clearly distinguish seven peaks for both curves acquired on the pristine film. Interestingly, the fundamental eigenmode at 13GHz does not yield the highest contribution and is significantly broader than the other peaks. A similar observation was also reported in the different context of picosecond ultrasonics on van der Waals nanolayers[38], suggesting a particularly efficient dissipation of the fundamental eigenmode towards the substrate compared to higher order modes. Because of its larger wavelength, we can hypothesise that the fundamental mode experiences a higher degree of elastic coupling between the nickel and $SiO_2$[39], leading to more efficient dissipation towards the substrate. As a result, the opto-acoustic response is mainly determined by the n=2,3,4 order eigenmodes with the dominance of the 3$^{rd}$ mode at ~40 GHz. On the other hand, the fundamental mode is clearly contributing to the spectra of the delaminated cavities, while all eigenmodes above 80GHz are completely suppressed. Overall, the acoustic spectrum of the response of the cavities is dominated by lower frequency modes compared to the film. This analysis is another way to see the different "broadness" and shape of the strain profile in the cavity compared to the film. The spectral position of these eigenmodes can be estimated by following a simple approach. The delaminated cavity can be modeled as a homogeneous nickel layer of thickness $h$ with free external faces, and due to the symmetry of the excitation, only the longitudinal modes can be excited. The eigenfrequencies satisfy the following relation[25]:

$$f_n = \frac{v_l}{2h} n, \quad n = 1,2,3, \ldots \qquad (1)$$

where $v_l$ is the longitudinal speed of sound. As for the film, it can be modelled as a nickel layer on top of a semi-infinite $SiO_2$ substrate. The nickel layer leaks out mechanical energy to the substrate, hence the eigenfrequencies $\tilde{f}$ are complex, their imaginary part representing the acoustic damping. Since the acoustic impedance of nickel ($Z_{Ni} = 5.37\ \mathrm{kg\,m^{-2}s^{-1}}$) exceeds the one of the $SiO_2$ ($Z_{SiO_2} = 1.31\ \mathrm{kg\,m^{-2}s^{-1}}$), the real part of the eigenfrequencies obeys to Equation (1) as well[26]. The position of the eigenfrequencies is highlighted in Figures 2c,d with blue (film on top of the $SiO_2$ substrate) and red (freestanding film) triangles. The eigenfrequencies estimated from Eq. (1) agree well with our experimental findings, confirming that the peaks shown in the Figures 2b correspond indeed to longitudinal acoustic modes.

To further support our discussion, we developed an opto-thermo-mechanical model and solved it with the finite element method (FEM) implemented in Comsol Multiphysics® (see Supplementary

Note 4 for further details)[27–29]. The pump pulse at 515 nm illuminates from the top the nickel layer (we assume a fluence equal to 5 mJ/cm$^2$ as in the experiments) and penetrates for less than 30 nm (see Supplementary Figure S5). This causes a localized increase of electronic temperature in the nickel layer close to the air side up to 800 K, just after the pump pulse has interacted with the system (see also Supplementary Figure S6). The electrons transfer energy to the phonon population, and hence a phononic temperature increase occurs. On the air side of the Ni layer, the phononic temperature reaches 150 K approximately 2 ps after pump excitation (see Supplementary Figure S6). After 2 ps, a thermal expansion of the first 30 nm of the nickel layer takes place, triggering the generation of an acoustic pulse, which travels towards the backside of the Ni layer and is reflected when it arrives at the second interface. The reflection is only partial in the case of the film on top of the substrate, whereas the totality of the mechanical energy is reflected, in principle, at the bottom boundary of the freestanding film. This can be appreciated in the simulations where for the freestanding we have an almost infinite number of echoes, while the echoes vanish quickly for the film on top of the $SiO_2$ substrate (see Supplementary Figure S7 and S8, respectively). After the reflection, a wavefront propagates towards the top part of the nickel layer and when it arrives in the first 30 nm of the layer – the zone interacting with the probe beam – a transient variation of the mechanical displacement (see Supplementary Note 4) due to the echo is recorded by the probe, changing the transient reflectance signal detected in our setup. The FFT transform of the longitudinal displacement $u_y$ calculated with FEM is reported in Figure 2c,d (continuous lines), showing good qualitative agreement with the peaks of observed in the FFT of the transient reflectivity traces and confirming the consistency of the model with the experimental data and observations. The minor discrepancies between the calculated eigenmodes and the experimental ones can be attributed to the idealized conditions assumed in our simulations, which are performed on an ideal system. In contrast with the experimental results, it is also worth mentioning that in our simulations the film shows an intense peak at the fundamental frequency. This discrepancy can also be explained by assuming that the experimental curves are affected by zero detection efficiencies at certain frequencies resulting from the zero-overlap integral between the acoustic eigenmode and the photo-acoustic sensitivity function[31]. Remarkably, the comparison between experimental and theoretical curves unambiguously shows that our simple model, where the delaminated cavity is treated as an ideal freestanding film, cannot fully capture the physics governing propagation of acoustic pulses. Grain boundaries and surface irregularities play an important role in shaping the opto-acoustic response of the delaminated cavities outside the simple reduction of attenuation.

Disentangling the dynamics associated of individual eigenmodes can provide valuable insight into our experimental findings. To achieve this, we calculate a short-time Fourier transform (STFT) for all measured pump-probe traces. The resulting spectrograms are shown in Figure 3. The two spectrograms associated to the pristine film (Figure 3a,b) reveal how the fundamental eigenmode is barely excited and its contribution vanishes completely after the first roundtrip. The excitation of the second eigenmode is also comparatively weak, while modes n=3,4,5 dominate the response. Regardless of the initial excitation amplitude, all modes in the pristine film undergo significant attenuation above 50% after each roundtrip. In contrast, the first two eigenmodes are efficiently excited in the delaminated cavities and their lifetime is considerably longer compared to the pristine film. Remarkably, in two of the cavities considered (cavity 3 and 4) the n=2 mode at 26GHz dominates the response and propagates essentially unperturbed over the time window we investigated. In agreement with our previous considerations, the spectrograms further highlight how the acoustic wave-packet propagating in the delaminated cavities lacks high-frequency modes >60GHz.

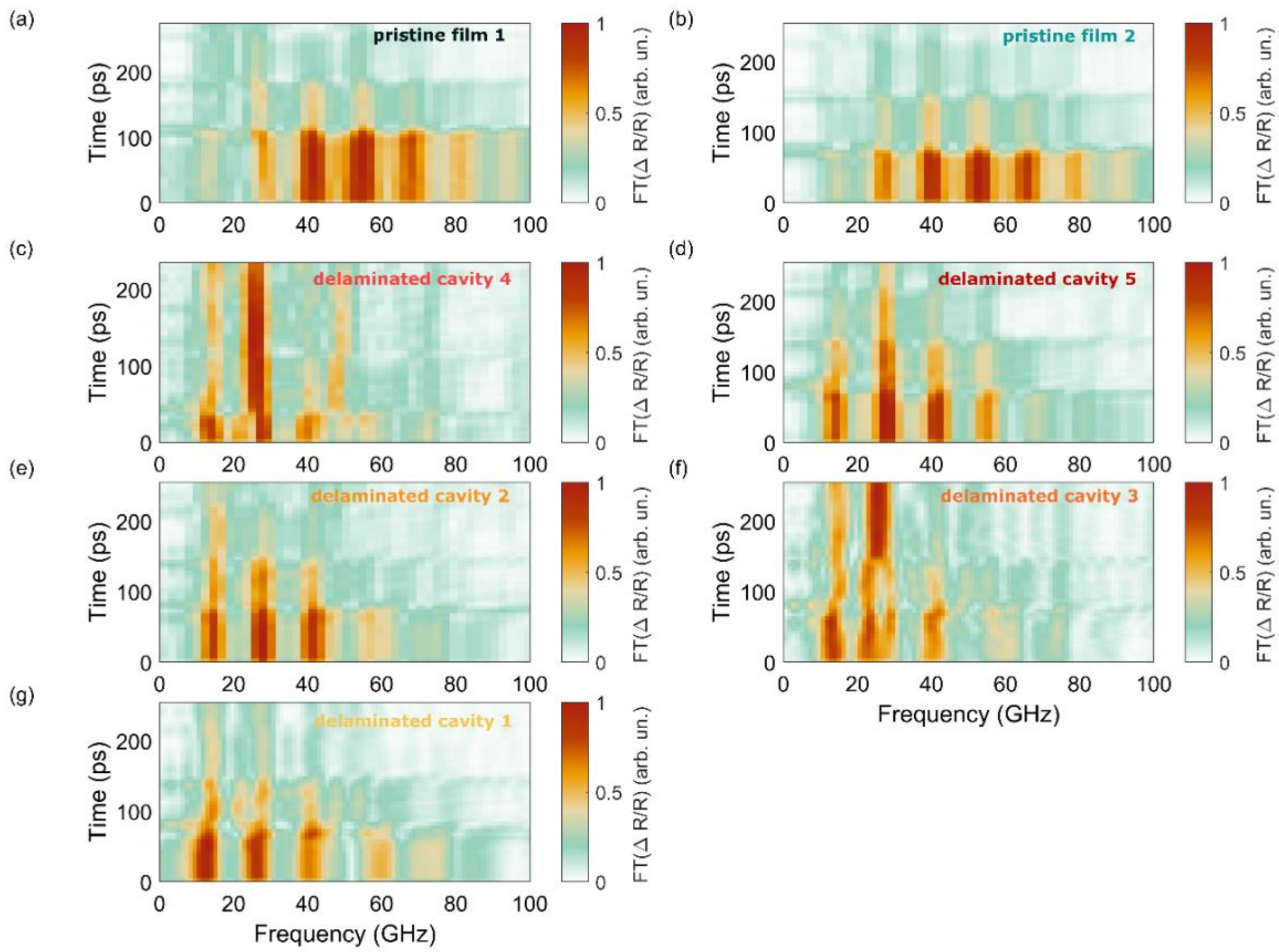

**Figure 3.** Propagation of individual acoustic eigenmodes for the pristine film (a,b) and for the different delaminated cavities (c-g). The spectrograms are obtained from the experimental pump-probe traces shown in Figure 2a using a short-time Fourier transform algorithm.

Comparison of the time evolution of the individual eigenmodes suggests that acoustic-pulse propagation in the pristine film and delaminated cavities is governed by different mechanisms.
To understand the dominant process in the two cases, we evaluate the damping associated to each eigenmode. To achieve this, we decompose the differential reflectance curves in a series of damped harmonic oscillators, each corresponding to an acoustic eigenmode of the pristine film/ delaminated cavity. In particular, we fit our experimental differential reflectance pump-probe traces with the function

$$\frac{\Delta R(t)}{R} = \sum_{n=1}^{N} A_n e^{-\gamma_n t} \cos(\omega_n t + \phi_n) \qquad (2)$$

where we consider N damped oscillators with amplitude $A_n$, decay rate $\gamma_n$ and eigenfrequencies $f_n = \omega_n/2\pi$. As indicated by the Fourier analysis discussed above, respectively N=5 and N=7 eigenmodes are needed to reproduce the response of the delaminated cavities and pristine film. The good agreement between the experimental data and best fit curves both in the time and frequency (solid curves in Figures 2a,b) domain validates the soundness of our methodology.
In Figure 4a we plot the extracted decay rates $\gamma_n$ as function of frequency $f_n$ for the eigenmodes considered. The error bars are obtained from the fitting procedure.

In Ref. [21] Ghita et al. applied the same analysis to the pump-probe measurements acquired by Kim et al[6] on a freestanding film obtained via dissolution in water. A quadratic dependence of the decay rate on frequency was observed up to 80GHz, indicating that that phonon-phonon scattering is the main attenuation mechanism. In our case, a fit of the decay rates to a power law $y = a + bx^c$ yields for both spots investigated on the pristine film an exponent $c \simeq 1.5$, in good agreement with the observations of Ghita et al. On the other hand, the measurements on the delaminated cavities yields a strikingly different $c = (4.6 \pm 0.8)$. This faster-than-quadratic dependence demonstrates that the damping of acoustic waves in the delaminated cavities is governed by a mechanism other than phonon-phonon scattering. On the other hand, a $\gamma \propto \omega^4$ dependence is expected when attenuation is dominated by Rayleigh-scattering[32,33]. In this context, the $c \simeq 4$ exponent strongly indicates that scattering due sub-acoustic-wavelength irregularities induced by the fabrication process (such as the larger crystallites and the irregularities at the bottom interface visible in the SEM images of Figure 1b) govern the high-frequency opto-acoustic response of the delaminated cavities.
Finally, in panel 4b we extract the quality factors $Q_n = \omega_n/2\gamma_n$ associated to the various eigenmodes and compare it with the typical quality factor $Q_G \simeq 12$ of magnon resonances in nickel[21]. As discussed by Ghita et al, the condition $Q_n \geq Q_G$ is a necessary requirement for efficient excitation of magnetization dynamics by the acoustic pulse. This condition is satisfied by the second and third eigenmodes in all cavities, demonstrating that delaminated cavities can host acoustic pulses suitable for control of magnetization dynamics[35] up to 60GHz. Above this frequency, we observed a decrease in quality factor. The frequency-scaling points towards phonon-defect scattering as the main attenuation mechanism, a behavior we attribute to structural changes in the nickel layer induced by the delamination process.
In conclusion, we performed an ultrafast spectroscopy study of acoustic phonons excited by femtosecond light pulses and confined in a cavity of a freestanding nickel layer fabricated using laser delamination. Combining a Fourier analysis with numerical multi-physics simulations, we showed that the freestanding cavity detached from the substrate displays a different acoustic response, with a significantly weaker damping of low-frequency modes compared to a reference thin film on a $SiO_2$ substrate.

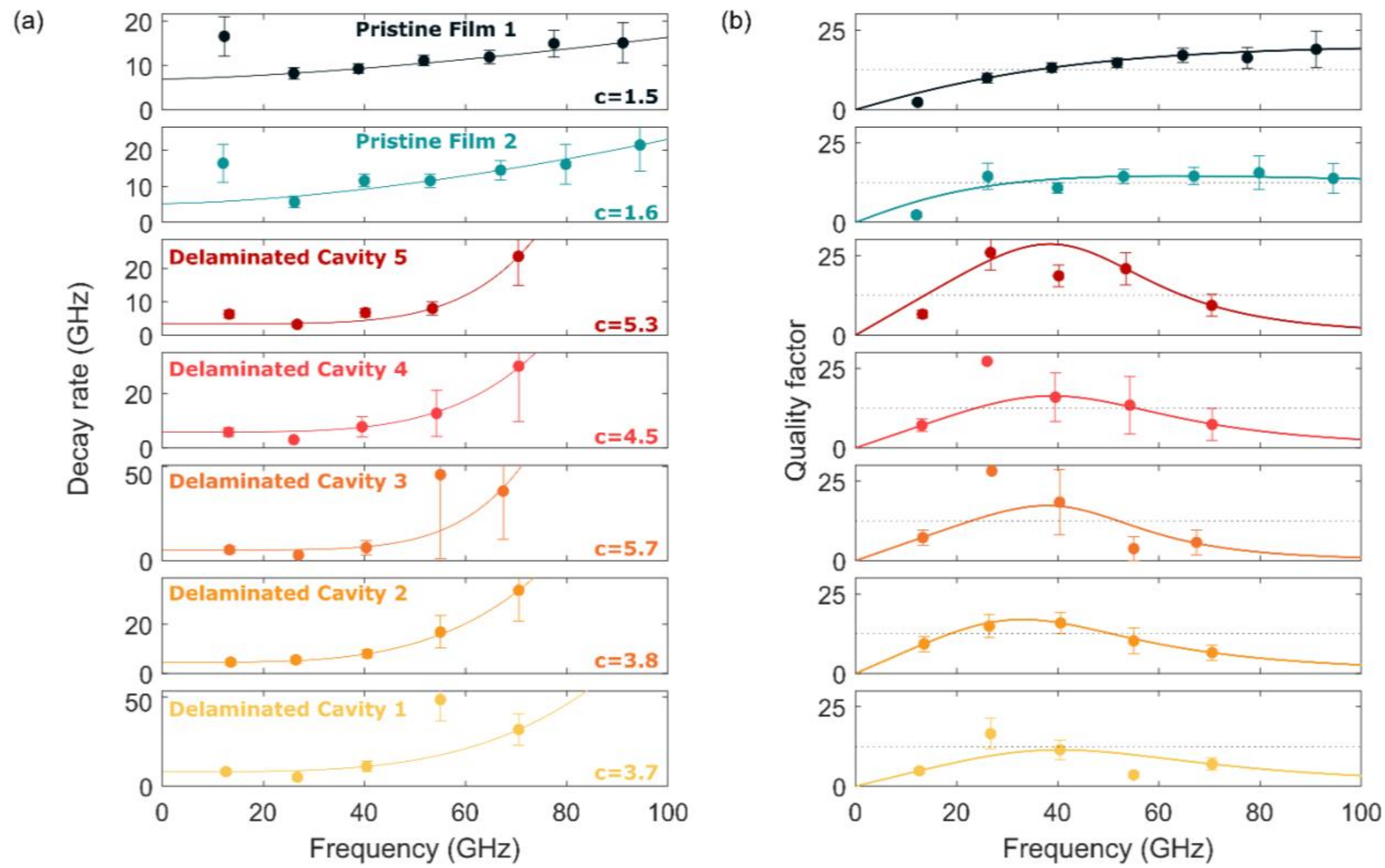

**Figure 4.** (a) Decay rates as a function of frequency obtained from the fit of the pump-probe data. A power law fit $y = a + bx^c$ is shown as a solid line as guide to the eye for all cases. (b) Q-factor associated to different eigenmodes. The solid lines are obtained from the fits of panel (a) and shown as guide to the eye. The black dashed line indicates the Q-factor of magnon resonances in nickel $Q_G \simeq 12$ as discussed in Ref. [19].

However, we also observed a surprisingly efficient suppression of higher-order modes in the delaminated cavities. To understand the origin of this behaviour, we quantitatively estimated the damping associated to the different eigenmodes of both delaminated cavities and pristine film. While the damping of the pristine film is compatible with the expected quadratic frequency-dependence, a fourth-order power law best describes the observations on the delaminated cavities. These findings, supported by SEM images, show that the delamination process causes the formation of defects in the nickel crystal, which ultimately limit the high-frequency response of freestanding cavities. Finally, we estimated the quality-factor associated to the different eigenmodes and compared it with that of the magnon resonances of nickel. Our analysis shows that delaminated cavities are indeed a promising platform for ultrafast modulation of acoustic waves. We showed that opto-acoustic control of low-frequency (<60GHz) magnetic excitations is already viable. While high-frequency (>100 GHz) operation remains a desirable goal for technological applications, our findings indicate that excitation of >60 GHz acoustic modes is currently inefficient in delaminated cavities. Crucially, we identified scattering between phonons and fabrication-induced defects as the limiting factor. In view of these findings, we envision that further optimization of the fabrication technique can overcome current limitations, paving the way for faster modulation of acoustic pulses in delaminated cavities.

## SUPPLEMENTARY MATERIAL

Details on fabrication of nickel cavities, sample preparation for SEM imaging, pump-probe measurements, numerical simulations, and Fourier analysis of the single echoes.

## ACKNOWLEDGEMENTS

N.M. acknowledges support from the Swedish Research Council (Grant No. 2021-05784), the Knut and Alice Wallenberg Foundation through the Wallenberg Academy Fellows Programme (Grant No. 2023.0089) and the European Research Council (ERC Starting Grant No. 101116253 'MagneticTWIST'). A.V.R., D.B. and N.M. acknowledge support from the European Commission (Grant No. 964363) and the Luxembourg National Research Fund (Grant No. C19/MS/13624497). D.B. acknowledges support from the European Regional Development Fund (Project 'UltrafastLux 2' Grant No. 2023-01-04). Y.U.E. and P.V. acknowledge support from the Spanish Ministry of Science and Innovation under the Maria de Maeztu Units of Excellence Program (Grant No. CEX2020-001038-M) and Project No. PID2021-123943NB-I00 (OPTOMETAMAG), as well as by Predoctoral Fellowship No. PRE2022-103017 and the Research Program for International Talents at Ecole Polytechnique. M.A.V. and M.G. acknowledge financial support of MUR "CARAMEL" project (CUP D73C24000220001).

# SUPPLEMENTARY MATERIAL

**Supplementary Note 1 – Fabrication of nickel cavities**

Freestanding nickel cavities were fabricated using a pulsed laser (EXPLA, 30ps @ 1064 nm, 10 Hz repetition rate) focused on 230 nm thick nickel thin film evaporated on top of a $SiO_2$ substrate using a 20 cm lens under an incident beam at 45 degrees impinging from the substrate side. The sample was scanned through the laser focus at a constant speed of 1 mm/s resulting in fabrication of lines of nearly identical structures separated by a distance on 100 micrometers (see Supplementary Figure S1*)*. Several lines of structures were fabricated at different pulse energies (in the μJ) range controlled by a combination of a motorized λ-half plate and a polarizer.

**Supplementary Note 2 – Sample preparation for SEM imaging**

In Figure 1b of the main text, we show cross-sections of the film and the cavity imaged by using scanning electron microscopy (SEM). To perform these measurements, we have performed precision ion beam milling of the structure using focused ion beam (FIB) followed by high resolution SEM imaging of structure cross sections. Prior to FIB-cutting, spallation cavities have been covered with a thin layer of Pt, initially using electron beam-induced deposition followed by a thicker Pt layer deposited by ion-beam induced deposition. This step is essential for preventing the surface of the nickel film from damage during the thick protective Pt-layer deposition. Another advantage of using e-beam deposited Pt layer is that it provides good contrast at the interface of Ni-film and ion-deposited Pt layer. Both protective layers allow to preserve the initial surface and prepare a clean and sharp FIB cut. The SEM images of the cross-sections, presented in Figure 1a in the main text, were obtained using a HeliosNanoLab 450S (ThermoFisher) at the accelerating voltage of 2–5 kV and the beam currentof 50-100 pA. From these images we can get information about the possible differences in the crystal structure after the laser spallation fabrication procedure.

**Supplementary Note 3 – Pump-probe measurements**

A first laser pulse (pump) is focused onto the sample and results in a thermally induced expansion that generates a time-dependent strain profile at the first interface (air/nickel) that travels back and forth within the nickel cavity/film, as sketched in Figure 1c. The propagation of this acoustic pulse can be studied using a second laser pulse (probe). Subsequent reflections of the acoustic pulse at the air/Ni interface are detected as a periodic modulation of the pump-probe transient reflectance signal ΔR/R. The measured transient reflectance, that is the pump-induced reflectance change, is defined as

$$\frac{\Delta R}{R} = \frac{R_t - R_0}{R_0}$$

where $R_t$ denotes the reflectance of the system after excitation and at a certain time delay between the pump and the probe pulses, while $R_0$ is the steady-state reflectance of the sample. The pump-probe setup, which is presented in Supplementary Figure S1, is based on an Yb:KGW laser system (Light Conversion®), delivering 220 fs pulses centered at 1030 nm with 500 mW average power at a repetition rate of 100 kHz. As illustrated in Supplementary Figure S1, a beam splitter (BS) divides the initial pulse train into two branches to form the pump and probe lines of our setup.

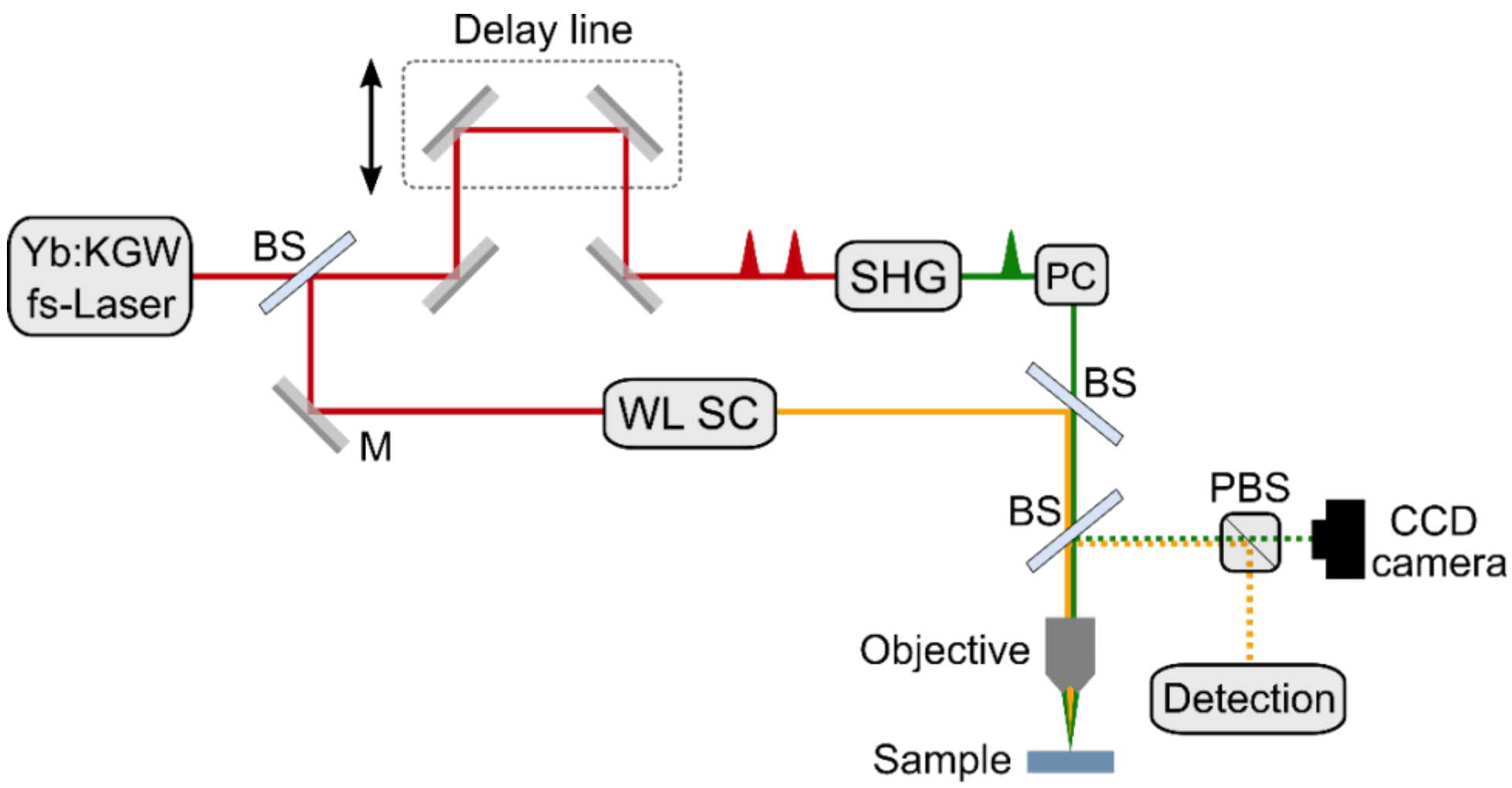

**Supplementary Figure S1.** Schematic representation of the setup used to carry out the experiments. The laser is divided into two lines, to respectively drive the pump (green) and the probe (yellow) beams towards the sample to perform the pump-probe spectroscopy measurements. A reflective objective is used to focus the pump and probe beams, as well as to image the sample onto a CCD camera. The reflected probe beam is filtered from the pump using a polarizing beam splitter (PBS) and detected by a photodiode. Acronyms: beam splitter (BS), mirror (M), white light supercontinuum (WL SC), Pockels cell (PC), second harmonic generation (SHG).

As pump beam, we use the second harmonic (SHG) of the laser fundamental (515 nm), modulated at half repetition rate, 50 kHz, using a Pockels cell (PC). The results shown here are performed with a pump fluence of 5 mJ/cm$^2$.

After excitation, the opto-mechanical dynamics is tracked by the probe beam, consisting of a white light supercontinuum (WLSC) generated using a YAG crystal and providing a broadband pulse from 500 to 950 nm with a duration of approximately 200 fs. The pump and probe beams are then focused onto the front size of the cavity at normal incidence by using a reflective objective (numerical aperture 0.5), to a spot size of approximately 20 μm for the pump beam and of 5 μm for the probe. Having a sufficiently small spot size is necessary to isolate the opto-acoustic response of a single cavity, which presents lateral dimensions in the range of 15 to 20 μm (see Supplementary Figure S2). The objective is also used to image the sample onto a CCD camera and verify that we are indeed pumping the center of the desired cavity structure (see Supplementary Figure S3). After interaction with the sample, the reflected probe beam is recollimated using the same objective, filtered from the pump using a polarizing beam splitter and detected by a photodiode. To minimize noise in the measurements and easily separate the signals from the two beams, pump and probe polarizations are set perpendicular to each other. Additional measurements at different fluence values are presented in Supplementary Figure S4.

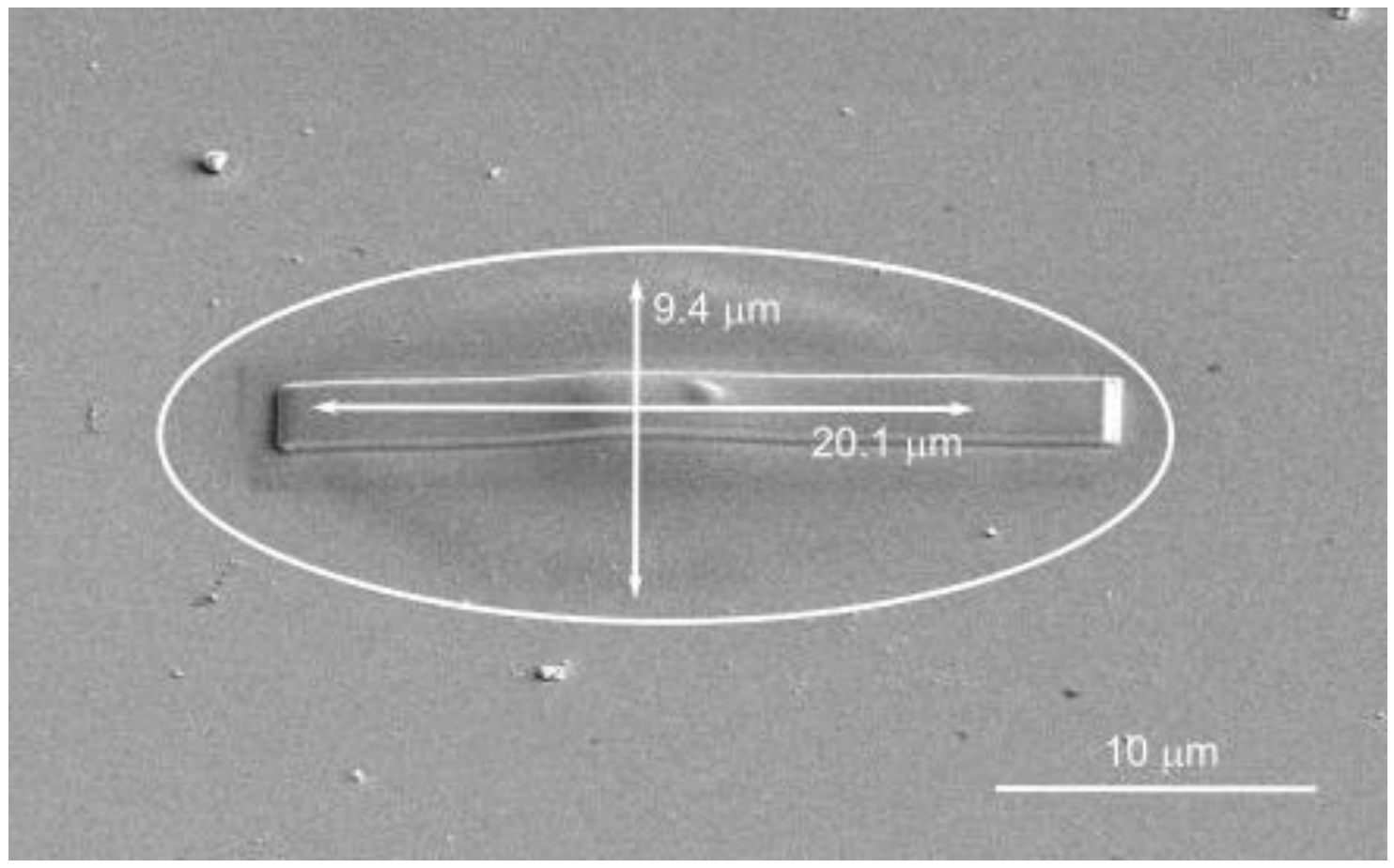

**Supplementary Figure S2.** Top-view SEM image of a nickel cavity, circled in white, indicating the lateral dimensions of the bubble. The rectangular structure on top is the Pt layer used to perform FIB and then SEM imaging of the vertical profile of the cavity.

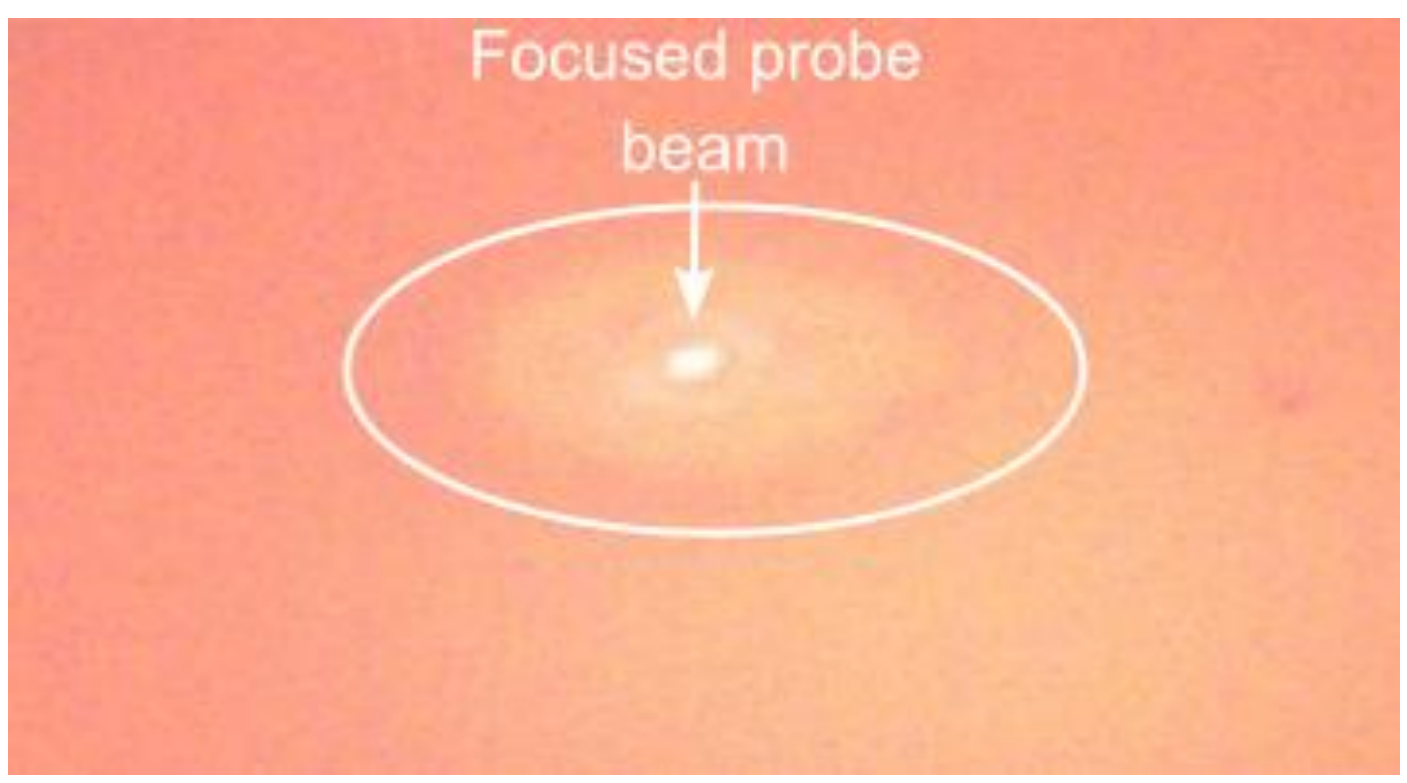

**Supplementary Figure S3.** Top-view of a cavity taken with the reflective objective use in the experimental setup with the probe beam at the center of the cavity. The circle represents the lateral dimensions of the cavity, proving that we are probing the center of the structure.

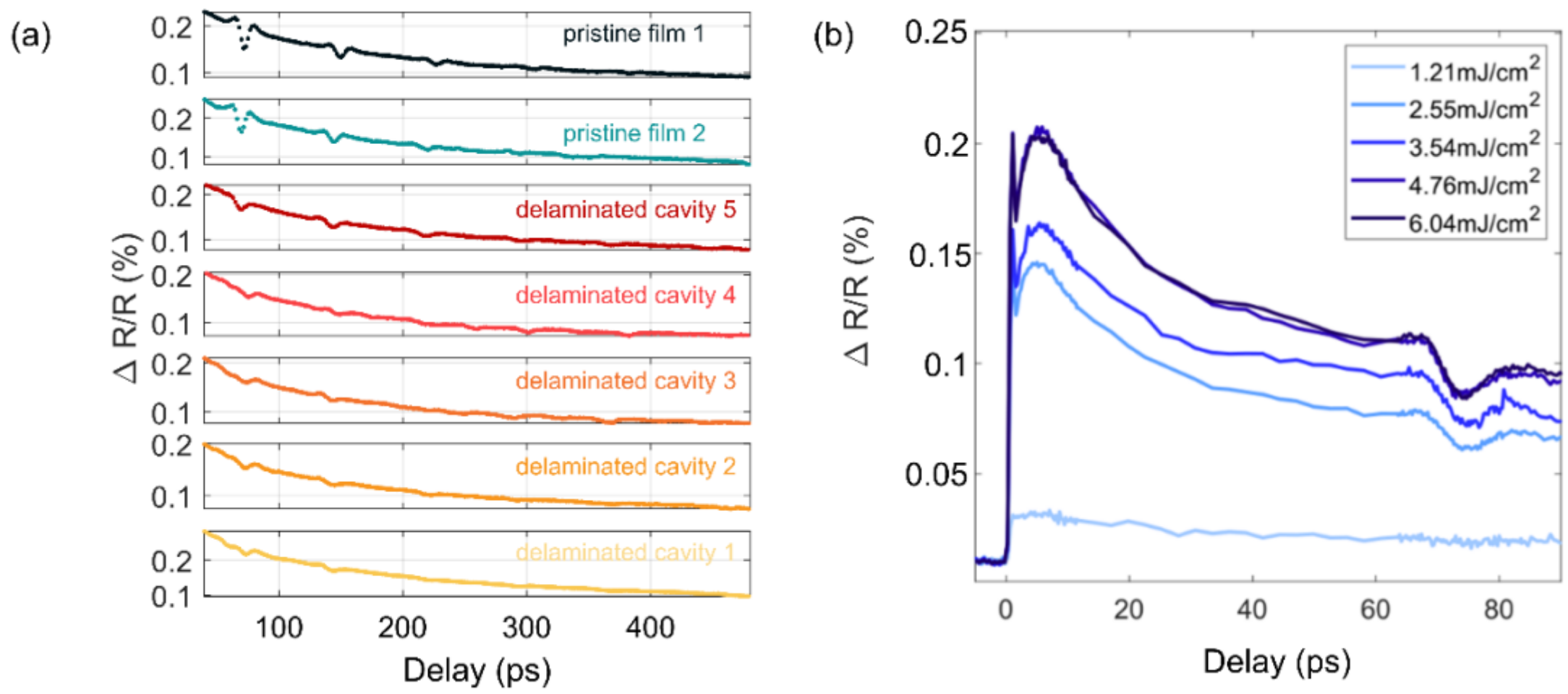

**Supplementary Figure S4.** (a) Pump-probe traces analyzed in the main text, before subtraction of the exponential baseline. (b) Pump-probe measurements at different fluence values. Saturation is reached for a fluence close to 5 mJ/cm$^2$.

**Supplementary Note 4 – Details on numerical simulations**

We have developed an opto-thermo-mechanical multi-physics model, that has been solved with Finite Elements Method (FEM) implemented in COMSOL Multiphysics®. The model solves three physical processes (optical, thermal and mechanical) sequentially. The solution of each step is used to compute the next one. The system under consideration is a Ni layer of thickness $h = 220$ nm, sandwiched among a semi-infinite air domain and a substrate (that may be either air – for the cavity – or glass – for the film case). Since in the experiment the pump beam size exceeds the one of the probe, the latter can be considered to measure the system in a region where the excitation intensity is maximum and rather homogeneous. Thus, it is reasonable to approximate the probed part of the system as a 1D system (along y coordinate – propagation of both the pump pulse and the acoustic longitudinal wave excited by the pump beam) since the dependence on the x and z coordinates is a constant. Within this approximation, the excitation can be well approximated by a plane wave propagating towards negative y-direction and whose intensity matches the pump beam peak intensity. The temporal period on one optical cycle is below 2 fs, way faster than the duration of the pulse. For this reason, it is enough to consider a frequency domain approach rather than a transient analysis for the optical part of our numerical analysis. To exploit the potentialities offered by COMSOL Multiphysics®, the aforementioned geometry has to be cast into a 2D geometry (x-y coordinates), see the inset of Supplementary Figure S5. By doing so, the translational invariance along the z-axis is automatically enforced. Furthermore, to model the translational invariance along x-direction, periodic boundary conditions (BC) are applied on the sides perpendicular to the x-axis for all the considered physics. With this approach, the choice of the domain width along x does not influence the results (even if we fixed the latter geometrical parameter to be as the light wavelength, for numerical convenience).

***Optical part.*** Three domains are considered, one for the air ($n = 1$), on top of the Ni layer of thickness $h$ (with $n = 1.85$ and $k = 3.4$ @515 nm[1]), anchored to a substrate (either $n = 1$ for the cavity or $n = 1.45$ for the film), see the inset of Supplementary Figure S5 for a sketch. The top air domain height is fixed to $h_{air}$ =300 nm, and it is truncated with a perfectly matched layer (PML) to approximate it as semi-infinite and avoid spurious reflections. The substrate domain height is $h_{sub}$ =300 nm as well, and it is truncated with a scattering BC. Despite the latter avoids spurious reflections to mimic a semi-infinite substrate, this choice is irrelevant because no light is transmitted to the substrate due to the high absorption of the Ni layer. As anticipated, the pump is modeled as a plane wave, and introduced with a periodic port. In the experiment, the pump wavelength, FWHM temporal duration, and fluence are $\lambda$ =515 nm, $\tau = 200$ fs and $\Phi = 5\ \mathrm{mJ/cm^2}$, respectively. Therefore, the intensity of the plane wave is as the peak intensity of the pump pulse (we refer to Ref. [2] for further details):

$$I_0 = \sqrt{\frac{4\ln(2)}{\pi}}\frac{\Phi}{\tau} \approx 23.5\frac{\mathrm{GW}}{\mathrm{cm^2}}. \qquad (S1)$$

The mesh size was set as 1/10 of the wavelength in each domain, whereas it was smaller than 4 nm in the Ni layer, to carefully follow the electric field exponential decay.

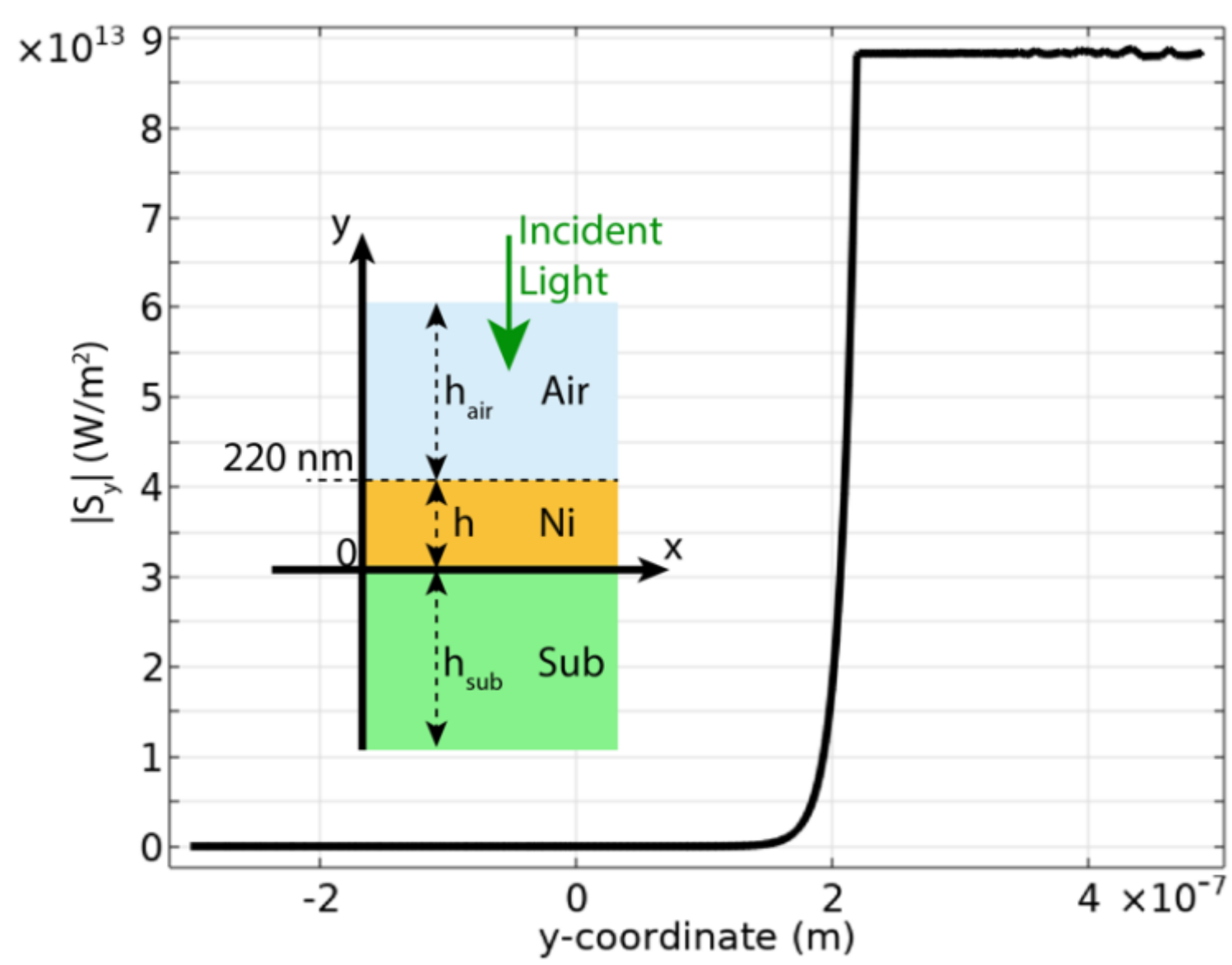

**Supplementary Figure S5.** Absolute value of the Poynting vector, y component, vs y. Inset: scheme of the 2D geometry used for the FEM.

The Maxwell's equations were solved with a frequency domain study (Electromagnetic Waves, Frequency Domain module was used), and the fields were calculated everywhere.

Hence, it is possible to calculate the dissipated power per unit volume due to Ohmic losses as:

$$Q_h(\boldsymbol{r}) = \frac{1}{2} Real(\boldsymbol{J} \cdot \boldsymbol{E}^*) = \frac{1}{2} \omega \varepsilon_0 Imag(\varepsilon_r) |E(\boldsymbol{r})|^2, \quad (S2)$$

where $\boldsymbol{J}$ is the total current density, $\boldsymbol{E}$ is the electric field, $\omega = 2\pi/\lambda$ and $\varepsilon_r$ the complex relative permittivity. Since the imaginary part of $\varepsilon_r$ is 0 for air and glass, losses are present in the Ni nickel domain only. The solution of the optical problem disclosed a light reflection of 62.4% at the air-Ni interface, whereas the remaining 37.6% is absorbed within the top part of the Ni layer. In Supplementary Figure S5 we report the Poynting vector as a function of y, showing an exponential decay within the first 30 nm (i.e. the skin depth) of the Ni layer. No transmission is recorded to the substrate, both for the cavity and the film. In accordance with the Poynting vector profile, the absorbed optical power is confined within the top part of the Ni layer.

***Thermal part.*** Once the optics has been solved, the thermal dynamics is considered. Since the pulse is 200 fs long, energy is delivered to the electronic population of the Ni layer, and only on later times energy is exchanged with the phonons. Hence a Two Temperature Model (TTM) is appropriate to describe this situation. The implemented equations are:

$$C_e \frac{\partial T_e}{\partial t} = \kappa_{T,e} \nabla^2 T_e + Q(\boldsymbol{r}, t) - G_{ep}(T_e - T_p), \quad (S3)$$

$$C_p \frac{\partial T_p}{\partial t} = \kappa_{T,p} \nabla^2 T_p + G_{ep}(T_e - T_p), \quad (S4)$$

where $T_e$ and $T_p$ are the electronic and phononic temperatures, respectively, $G_{ep}$ is the electron-phonon coupling coefficient, $\kappa_{T,e}$ and $\kappa_{T,p}$ are the electronic and phononic thermal conductivities, respectively, $C_e$ and $C_p$ are the electronic and phononic specific heat per unit volume, respectively. The system was initially at rest, so $T_e = T_p = T_0 = 293.15$ K everywhere. At later times, the pump pulse excites the system. The temporal duration of the excitation, not considered for the optics analysis, is now introduced by assigning a volumetric heat source term with a Gaussian temporal shape:

$$Q(\boldsymbol{r}, t) = Q_h(\boldsymbol{r}) \exp\left[-4 \ln(2) \frac{(t - t_0)^2}{\tau^2}\right], \quad (S5)$$

where $Q_h(\boldsymbol{r})$ was calculated previously with the optics step, $\tau$ is the pulse temporal FWHM, and the time maximizing the light intensity is $t_0 = 10\tau$ to avoid conflicts with the initial conditions.

For the cavity, equations S4 and S5 have been solved only within the 220 nm thick Ni layer, the presence of the air being neglected. On the Ni-air surfaces, adiabatic BCs were imposed for both the temperatures. No free electrons are available in the air, so they cannot carry any energy flux. As for the phonons, the air thermal conductivity is quite low, of the order of (or lower than) $10^{-2}$ W $\mathrm{m}^{-1}\mathrm{K}^{-1}$, and on the considered time scales it does not play. However, to be on the safe side, we have also tried to replace the adiabatic BCs for the phononic heat flux with radiative and convective BCs, but no significant changes on the obtained temperature profile were observed.
As for the film, we have also considered the $SiO_2$ substrate anchored to the bottom side of the Ni layer. The substrate is insulator, with no free electrons available. Hence, Eq. S3 was solved only in the Ni layer. Adiabatic BCs for the electronic heat flux were applied to both the Ni layer boundaries perpendicular to y. On the other hand, heat transfer to the substrate through phonons may occur, so Eq. S4 was solved also for the substrate. On the external boundaries perpendicular to y coordinate adiabatic BCs for the phononic heat flux were enforced. The substrate height is $h_{sub} = 8$ µm, hence in the time range of interest (1 ns after the arrival of the pump pulse), the thermal dynamics does not reach the end of the $SiO_2$ domain. As a result, the temperature profile obtained is equivalent to that of an infinite substrate, and the choice of the BC on the lower boundary of the substrate does not alter the result. The mesh adopted for the optics was replicated in the Ni layer and its proximal substrate. In the bottom part of the substrate, where no interesting dynamics occurs, the mesh was progressively released to reduce the numerical burden. As for the temporal step, to follow the pulse carefully, a temporal step of 10 fs was used. For times longer than 10 ps, the temporal step was released to 1 ps, to reduce the computational burden. The used thermal properties are summarized in Supplementary Table ST1. Eqs. S3 and S4 were solved together at the same time, with a time dependent study. Two coupled Heat transfer in Solid physics were implemented in COMSOL Multiphysics. The result consists in the two-temperature spatial and temporal profiles.

**Supplementary Table ST1.** Thermal parameters used in our analysis. The parameters for Ni are taken from Ref. [3]. Despite the precise value of these parameters may vary depending on the sample, they are somehow consistent also with what reported in Refs. [4,5].

| Thermal parameter | Value |
|---|---|
| $\kappa_{T,e}$ in Ni | $(0.2\ \mathrm{W\,m^{-1}K^{-2}}) \times T_e$ |
| $C_e$ in Ni | $\gamma_e T_e = (1077\ \mathrm{J\,m^{-3}K^{-2}}) \times T_e$ |
| $\kappa_{T,p}$ in Ni | $20\ \mathrm{W\,m^{-1}K^{-1}}$ |
| $C_p$ in Ni | $3.63 \times 10^6\ \mathrm{J\,m^{-3}K^{-1}}$ |
| $G_{ep}$ in Ni | $8.55 \times 10^{17}\ \mathrm{W\,m^{-3}K^{-1}}$ |
| $\kappa_{T,p}$ in $SiO_2$ | $0.8\ \mathrm{W\,m^{-1}K^{-1}}$ |
| $C_p$ in $SiO_2$ | $1.76 \times 10^6\ \mathrm{J\,m^{-3}K^{-1}}$ |

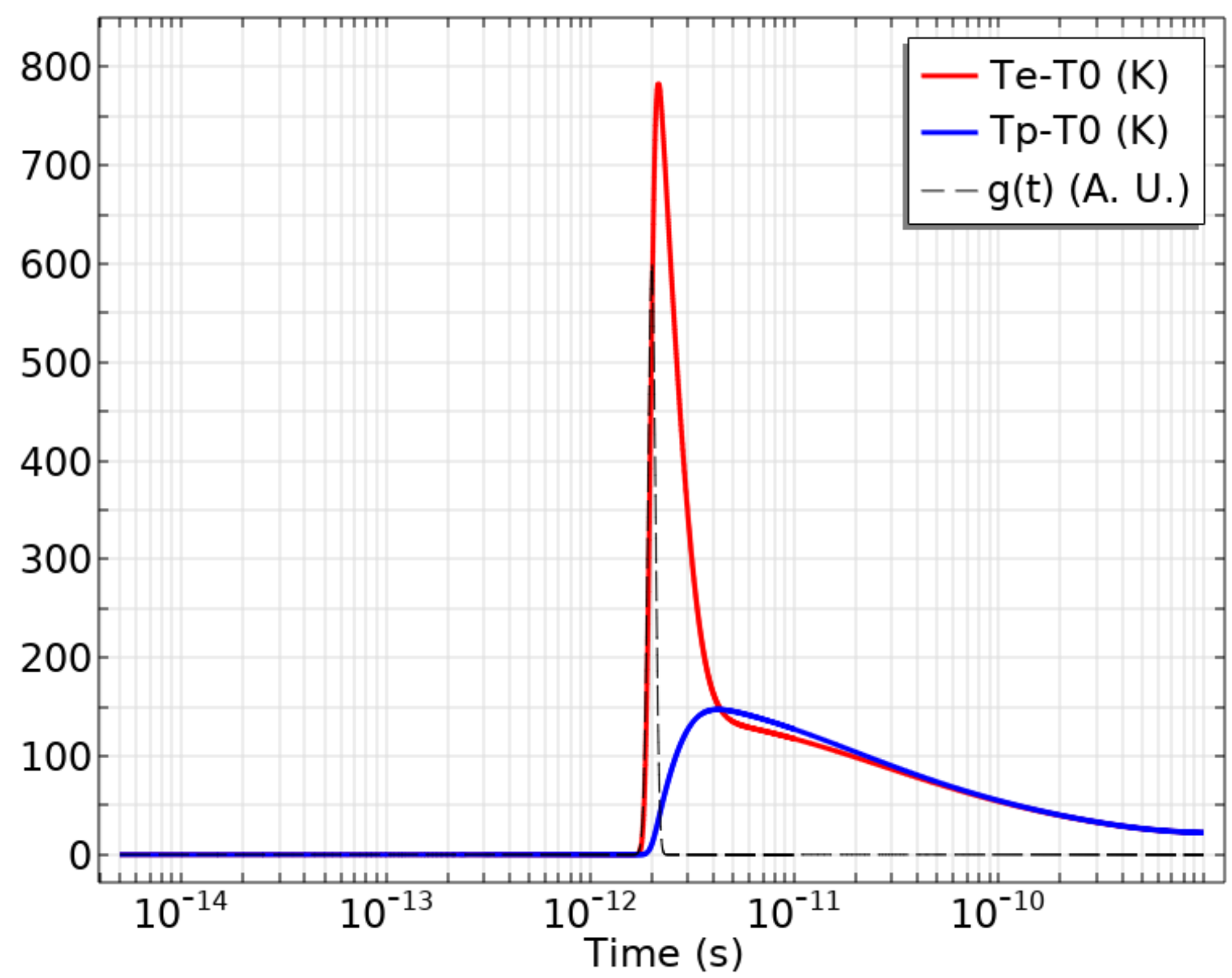

**Supplementary Figure S6.** Electronic (red) and phononic (blue) temperature increase vs time (horizontal axis, log scale). The temperatures were calculated at the coordinate $y = h - 10$ nm. The temporal shape of the pulse (black dashed line) has been reported for the sake of comparision.

In Supplementary Figure S6 we report the electronic and phononic temperature vs time in proximity of the top surface of the Ni layer, for the cavity. Since the thermal conductivity of the glass is very low, the heat flow to the substrate is very inefficient and the temperature profile in Ni layer in the simulated time range for the film case is very similar to the one obtained for the cavity. At 1 ns, the temperature increase in the top part of the substrate is as low as 20 K.

***Acoustic part.*** For this step we consider the same geometry, mesh and temporal discretization already described for the thermics. The acoustics rely on the solution of Navier's equation:

$$\rho \frac{\partial^2 \boldsymbol{u}}{\partial t^2} = \nabla \cdot \sigma, \qquad (S6)$$

$$\sigma = C\colon\left(\varepsilon - \alpha \Delta T_p I\right), \qquad (S7)$$

where $\boldsymbol{u}$, $\sigma$, $C$, $\varepsilon$, $\rho$, $\alpha$ and $I$ are the displacement, the stress, the stiffness tensor, the strain, the mass density, the linear expansion coefficient, and the identity matrix, respectively (we refer to Ref. 2 for further details). $\Delta T_p$ is the phononic temperature increase calculated previously with the thermics. The initial conditions are rest, i.e., $\boldsymbol{u}(t = 0) = \boldsymbol{0}$ and $\frac{\partial \boldsymbol{u}}{\partial t}(t = 0) = \boldsymbol{0}$. We enforced free boundary conditions on both the external boundaries perpendicular to y. In the case of the film, the substrate is long enough so that the lower boundary does not influence the result in the computed time interval, as if the substrate were semi-infinite. The parameters used for this computation are summarized in Supplementary Table ST2. The acoustics has been modeled in COMSOL

Multiphysics® with a Solid Mechanics module and solved with a time dependent study. Due to the symmetry of the excitation and of the geometry, only the longitudinal component of the displacement ($u_y$) is going to play. The displacement as a function of time, calculated in proximity of the Ni layer top surface, is reported in Supplementary Figures S7 and S8 for the cavity and the film, respectively. As we may see, in both the cases we see several displacement dips, each one arriving ~73 ps after the previous. These dips correspond to the acoustic echoes. For the case of the film, the amplitude of the dips decreases with time, a signature of the extrinsic damping related to the mechanical energy transfer to the substrate occurring along with every reflection at the lower boundary of the Ni layer. On the other hand, in the case of the cavity, the dips amplitude mainly remains constant with time, since no other damping mechanisms have been introduced in the model.

**Supplementary Table ST2.** Acoustic parameters used in our analysis. We used the parameters of Polycrystal Ni and fused silica (for the substrate), that are isotropic materials. The stiffness tensor components and mass densities are taken from Ref. [6]. The thermal expansion coefficients are taken from Refs. [7,8] for Ni, and Refs. [9,10] for fused silica.

| Acoustic parameter | Polycrystal Ni | $SiO_2$ |
|---|---|---|
| $C_{11}$ [GPa] | 324.0 | 78.5 |
| $C_{44}$ [GPa] | 80.0 | 31.2 |
| $\rho$ [kg m$^{-3}$] | 8900 | 2200 |
| $\alpha$ [K$^{-1}$] | $15.0 \times 10^{-6}$ | $0.5 \times 10^{-6}$ |

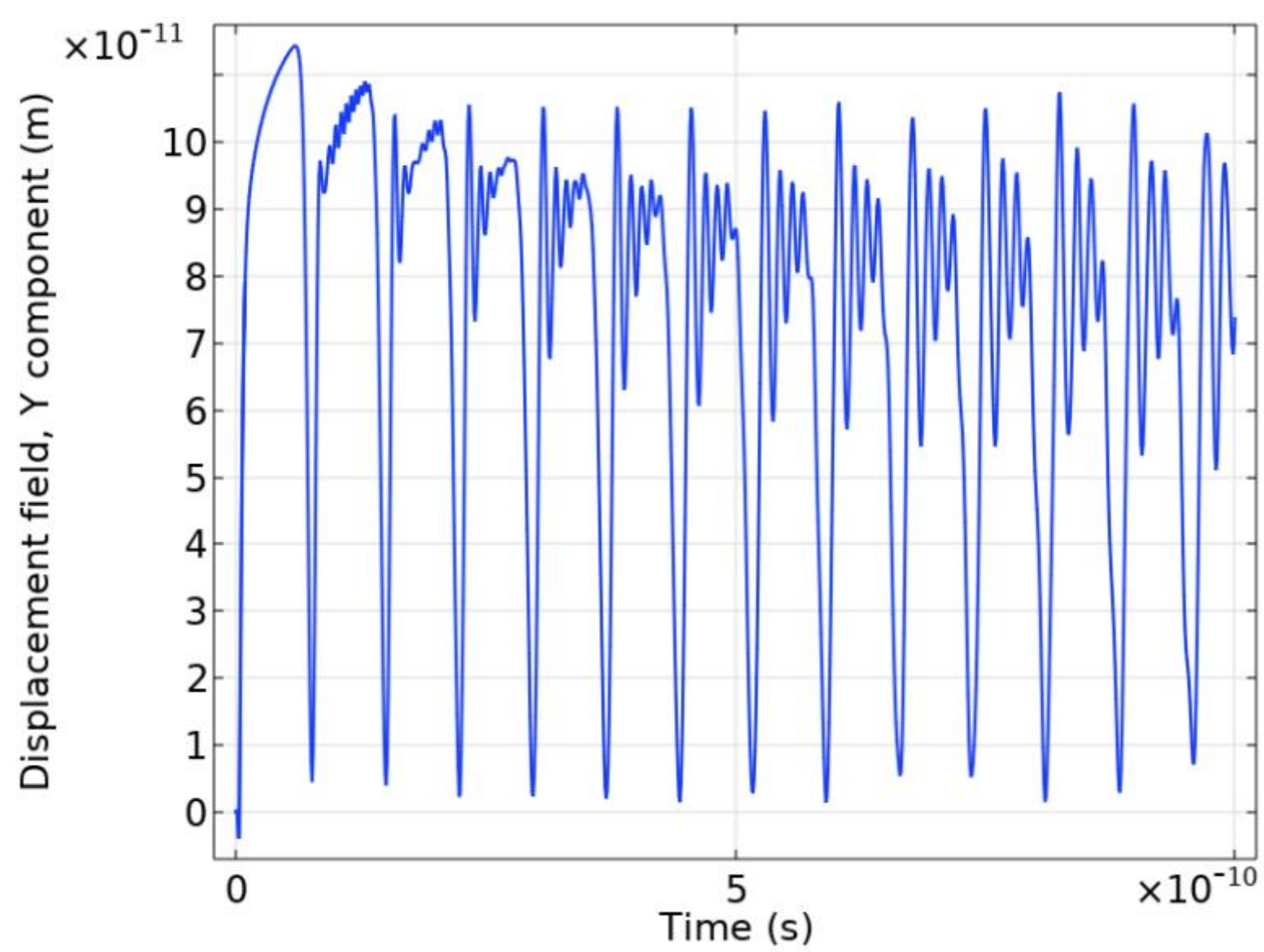

**Supplementary Figure S7.** Displacement $u_y$ as a function of time for the cavity. The displacement was calculated at the coordinate $y = h - 10$ nm from the nickel/air interface.

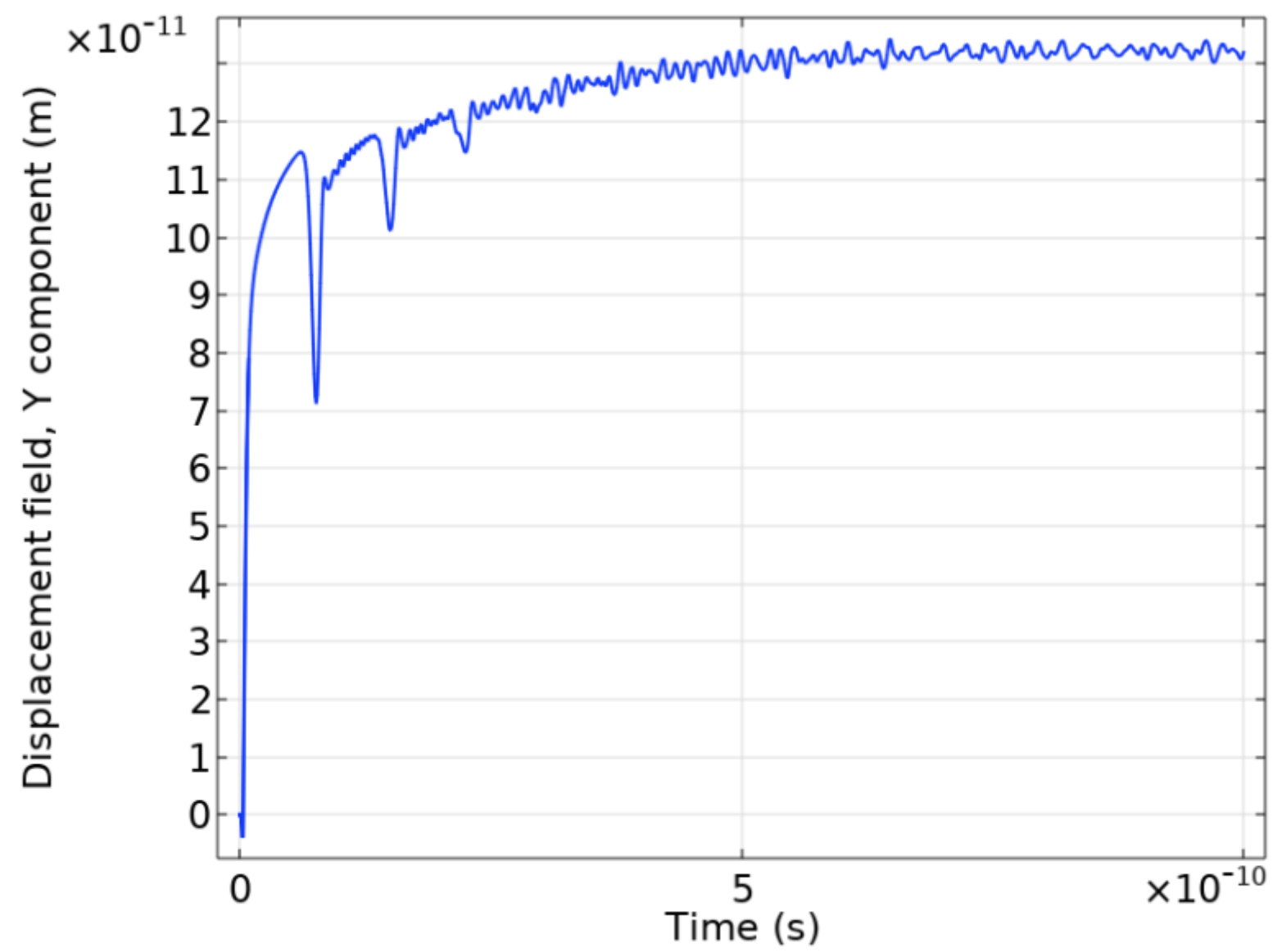

**Supplementary Figure S8.** Displacement $u_y$ as a function of time for the film. The displacement was calculated at the coordinate $y = h - 10$ nm from the nickel/air interface.

We pinpoint the fact that for both the cavity and the film the displacement is expected to be periodic in time as it can be represented as a superposition of the fundamental mode with a period $T_1 = \left(\frac{2h}{v_l}\right)$ and its higher harmonics with periods $T_n = T_1/n$:

$$u_y(y,t) = \sum_{n=1}^{+\infty} A_n(y) \cos\left[2\pi \left(\frac{v_l}{2h}\right) nt + \phi_n\right].$$

However, in Figure S8 we plot the displacement in the first ns after excitation. If we look at Figure S6, it is clear than within 1 ns the system is not yet thermalized, and so we expect the echoes to become "perfectly" periodic only after 1 ns, approximately after 10 ns after pump pulse excitation. Our intention was to show that in the case of a suspended cavity the number of echoes is much larger (infinite in the ideal case) compared to the case where the substrate takes part of the energy of the acoustic envelope every roundtrip, and so the number of echoes is much less (see Figure S7).

**Supplementary Note 5 – Fourier analysis of the single echoes**

In this note we propose an alternative analysis of how the damping can be obtained, which was inspired by the work of Zhu et al.[11]. As dataset for this analysis, we considered delaminated cavity 5 and pristine film 2. In Supplementary Figures S9a and S10a we plot the FFT of full temporal traces together with the FFT of the isolated first five echoes of both pristine film and delaminated cavity. The individual echoes are isolated by setting a time interval centered around the minimum peak of each echo. The time interval was varied until we obtained a stable FFT for each echo. The resulting FFT spectra are normalized to the sum of the FFT of all the selected echoes. We have extracted the points cut by the same frequencies for the six film eigenmodes and the five cavity ones (corresponding to the eigenfrequencies calculated in the main text and represented in Figure 2c,d), and fitted them with a first order exponential. From this fit we obtained the damping, with associated error bar, for each eigenmode. In Supplementary Figure S11 we compare damping and Q-factors extracted from the analysis of the individual echoes. We fit the damping curves with a power law $y = a + bx^c$ and obtain $c \simeq 2.2$ and $c \simeq 3.7$ for pristine film and delaminated cavity respectively, in good agreement with the results shown in the main text.

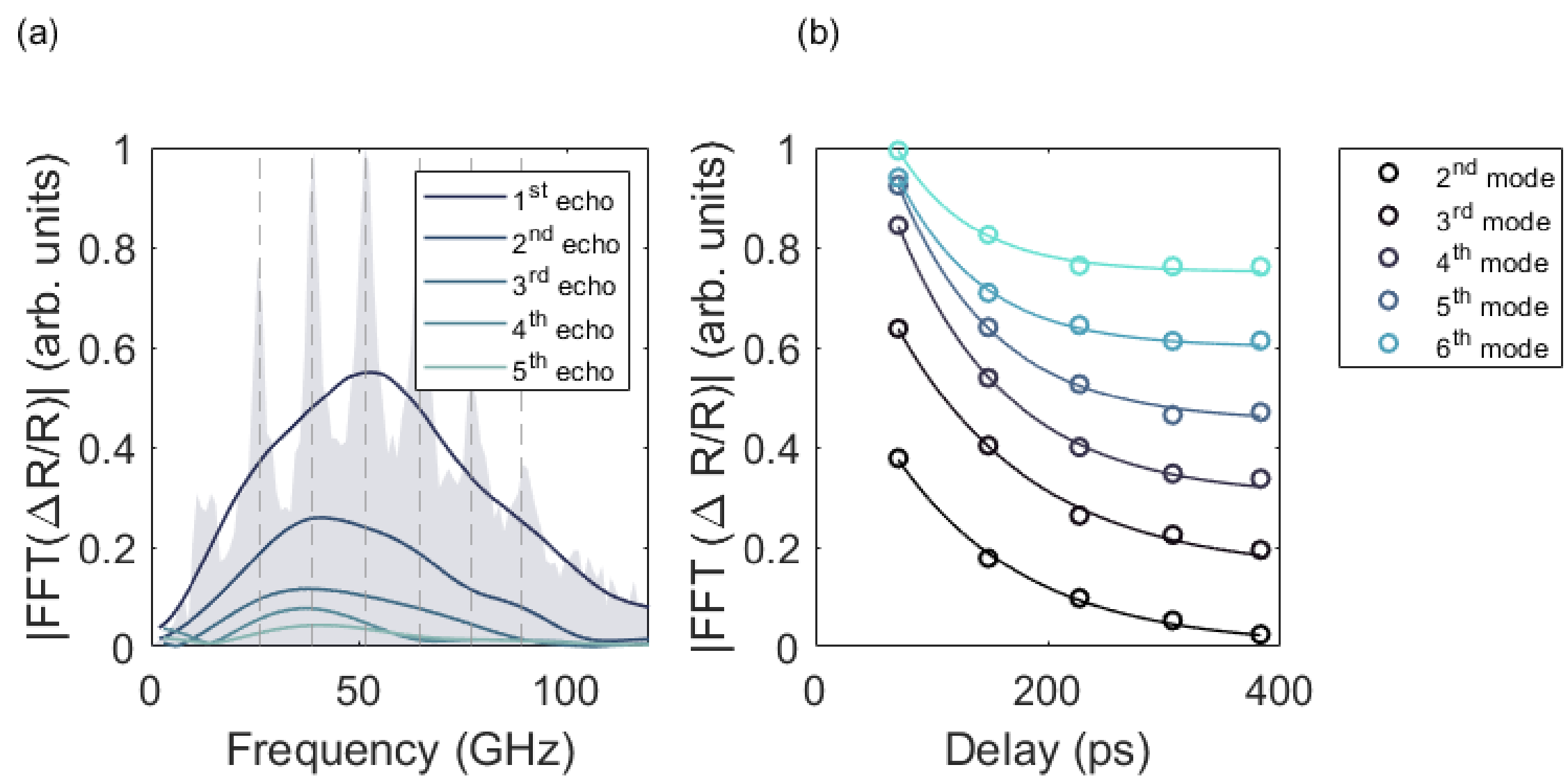

**Suppplementary Figure S9.** (a) FFT of the temporal trace (blue circles, Figure 3a) and the first five echoes for the film (dark to light blue solid lines). (b) Amplitude associated to each different eigenmode as a function of the time delay. The continuous line is the exponential fit from which we extract damping of the eigenmodes. The curves are offset for visibility

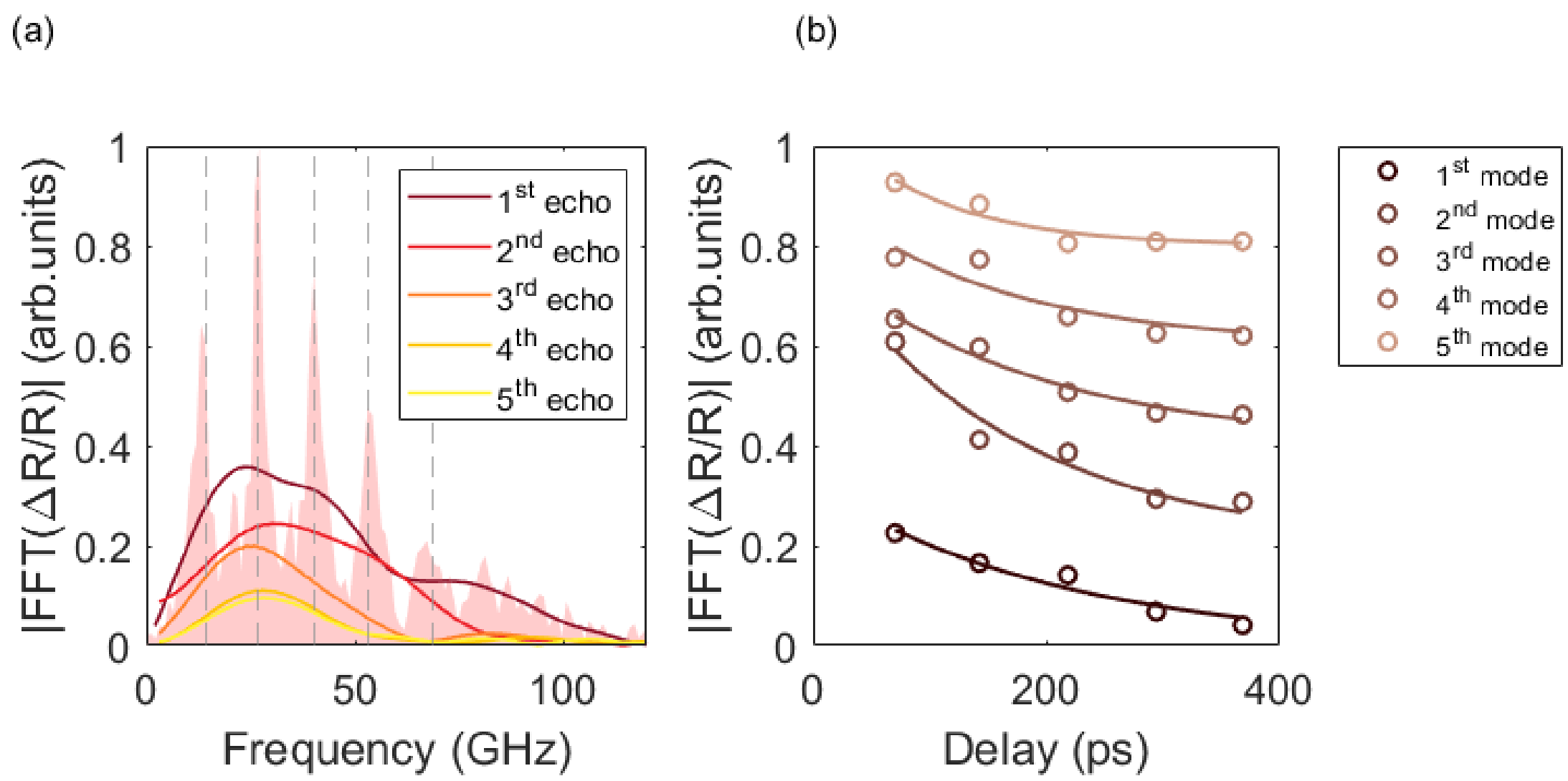

**Supplementary Figure S10.** (a) FFT of the temporal trace (Figure 3b) and the first five echoes for the cavity (dark red to yellow solid lines). (b) Amplitude of the FFT of the first four echoes at different frequencies as a function of the time delay. The continuous line is the exponential fit from which we extract the damping of the eigenmodes. The curves are offset for visibility.

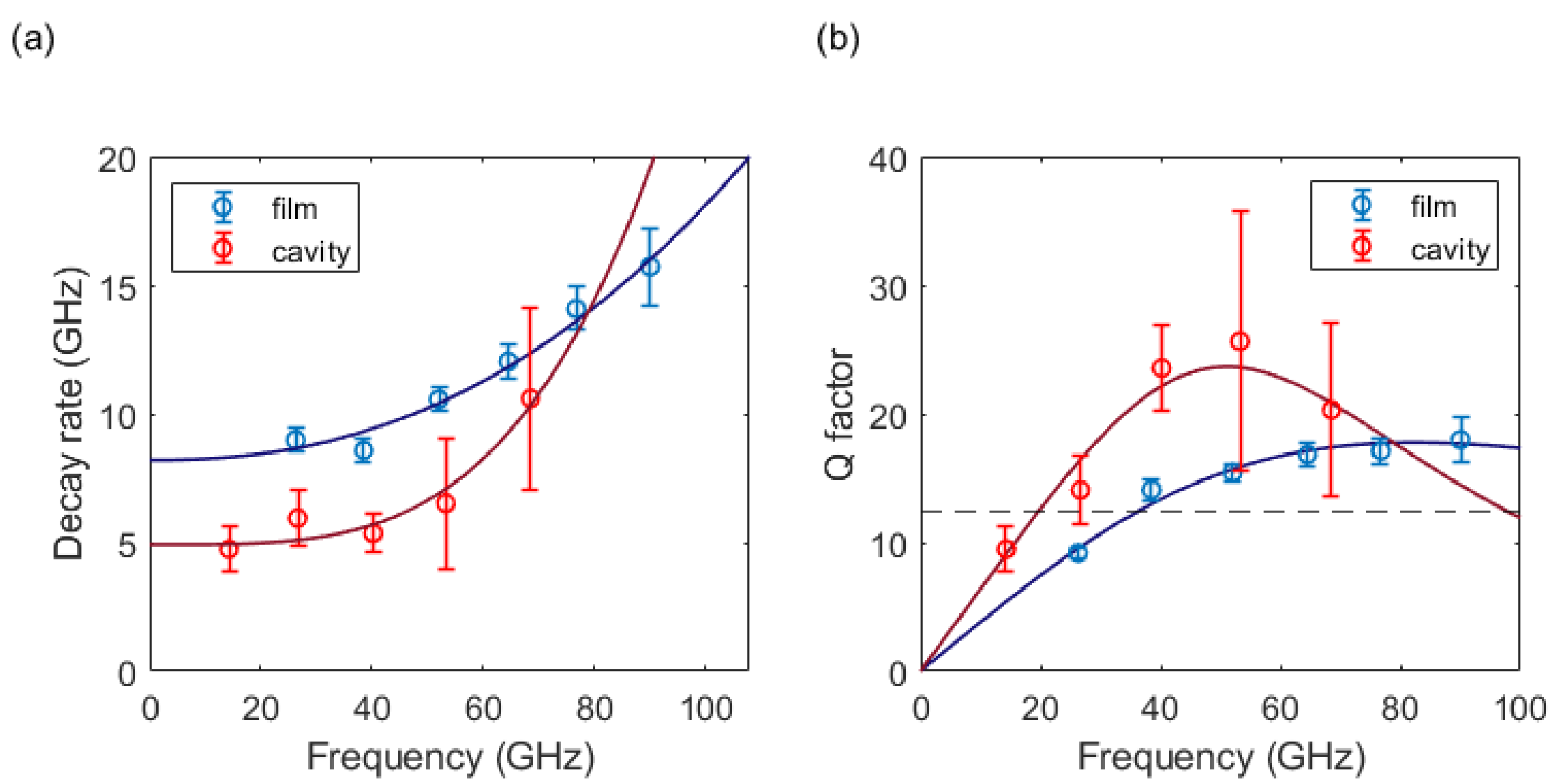

**Supplementary Figure S11.** (a) Damping of the individual eigenmodes for film (blue dots) and cavity (red dots), extracted from the analysis of the individual echoes . The solid line shows a power law fit $y = a + bx^c$. b) Experimental Q-factors (dots) and fit (solid line) extracted from the data in panel (a).

## Supplementary References